\documentclass[12pt]{iopart}

\usepackage{iopams}
\expandafter\let\csname equation*\endcsname\relax
\expandafter\let\csname endequation*\endcsname\relax
\usepackage{amsmath}
\usepackage{amssymb}
\usepackage{amsfonts}
\usepackage{bbold}
\usepackage{mathptmx}
\usepackage{psfrag,graphicx}
\usepackage{dcolumn}
\usepackage{bm}
\usepackage{color}
\usepackage{latexsym}
\usepackage{epstopdf}
\usepackage{color}
\usepackage[english]{babel}
\usepackage{latexsym}
\usepackage{bm}
\usepackage{epstopdf}
\usepackage{appendix}
\usepackage[T1]{fontenc}
\usepackage{subfig}

\usepackage{amssymb}

\definecolor{mygrey}{gray}{0.35}
\definecolor{myblue}{rgb}{0.2,0.2,0.8}
\definecolor{myzard}{cmyk}{0,0,0.05,0}
\definecolor{mywhite}{rgb}{1,1,1}
\definecolor{mywhite}{rgb}{1,1,1}
\definecolor{myred}{rgb}{1,0.,0.3}

\usepackage[colorlinks=true,citecolor=myblue,linkcolor=myred]{hyperref}

\newcommand{\ket}[1]{|{#1}\rangle}                       
\newcommand{\bra}[1]{\langle {#1}|}                      
\newcommand{\average}[1]{\langle {#1} \rangle}           

\newcommand{\Ignore}[1]{ }

\begin{document}

 \title[Dynamics of a harmonic oscillator coupled with a Glauber amplifier]{Dynamics of a harmonic oscillator coupled with a Glauber amplifier}

\author{R Grimaudo$^{1,2}$, V I Man'ko$^{3,4,5}$, M A Man'ko$^{3,4}$ and A Messina$^{2,6}$  }

\address{$^1$ Dipartimento di Fisica e Chimica dell'Universit\`a di Palermo, Via Archirafi, 36, I-90123 Palermo, Italy}
\address{$^2$ INFN, Sezione di Catania, I-95123 Catania, Italy}
\address{$^3$ P. N. Lebedev Physical Institute, Russian Academy of Sciences Leninskii Prospect 53, Moscow 119991, Russia}
\address{$^4$ Moscow Institute of Physics and Technology, Institutskii Per. 9, Dolgoprudny Moscow Region 141700, Russia}
\address{$^5$ Department of Physics, Tomsk State University, Lenin Avenue 36, Tomsk 634050, Russia}
\address{$^6$ Dipartimento di Matematica ed Informatica dell'Universit\`a di Palermo, Via Archirafi, 34, I-90123 Palermo, Italy}
\ead{roberto.grimaudo01$@$unipa.it}
\vspace{10pt}
\begin{indented}
\item[]February 2018
\end{indented}

\begin{abstract}
A system of a quantum harmonic oscillator bi-linearly coupled with a Glauber amplifier is analysed considering a time-dependent Hamiltonian model.
The Hilbert space of this system may be exactly subdivided into invariant finite dimensional subspaces. 
Resorting to the Jordan-Schwinger map, the dynamical problem within each invariant subspace may be traced back to an effective SU(2) Hamiltonian model expressed in terms of spin variables only.
This circumstance allows to analytically solve the dynamical problem and thus to study the exact dynamics of the oscillator-amplifier system under specific time-dependent scenarios.
Peculiar physical effects are brought to light by comparing the dynamics of such a system with that of two interacting standard oscillators.
\end{abstract}

%
\vspace{2pc}

\submitto{Physica Scripta}
%
%
%


\section{Introduction}

In 1982 Glauber introduced the idea of quantum amplifier \cite{Glauber0,Glauber} modelled through an inverted harmonic oscillator, that is, a system whose Hamiltonian may be represented as
\begin{equation}
-\hbar\omega(\hat{c}^\dagger \hat{c}+1/2),
\end{equation}
where $\hat{c}$ and $\hat{c}^\dagger$ are bosonic operator satisfying the usual commutation rule $[c,c^\dagger]=1$.
The eigenstates of the Hamiltonian are \cite{Tarzi}
\begin{equation}
\ket{n}={(\hat{c}^\dagger)^n \over \sqrt{n!}}\ket{0}
\end{equation}
where the null state is defined as $\hat{c}\ket{0}=0$.
It is important to point out that such a state is not the ground state of the system since the inverted oscillator does not possess a ground state being its spectrum unbounded from below.
Moreover, it is worth noticing that the operator $\hat{c}^\dagger$, still increasing the number $n$ of excitations, moves indeed the inverted oscillator system towards lower energy states (and vice versa for $\hat{c}$).

In Refs. \cite{Glauber0,Glauber} Glauber studies the thermodynamics of the amplifier system when it interacts with a bath of standard quantum harmonic oscillators.
The same system is investigated in Ref. \cite{Tarzi}.
In order to take into account interaction terms preserving the energy of the system $H_0=\sum_i\hbar\omega_i\hat{a}_i^\dagger\hat{a}_i-\hbar\omega_c\hat{c}^\dagger\hat{c}$, Glauber first considers the following model: $\sum_i\hbar\omega_i\hat{a}_i^\dagger\hat{a}_i-\hbar\omega_c\hat{c}^\dagger\hat{c}+\sum_i(k_i\hat{a}_i\hat{c}+k_i^*\hat{a}_i^\dagger \hat{c}^\dagger)$, where $\hat{a}_i$ and $\hat{a}_i^\dagger$ are the bosonic operators of the $i$-th standard oscillator.
Glauber makes evident the peculiar features of such a `non-standard' system by making a comparison with the more familiar system comprising a `standard' harmonic oscillator interacting with the bath.

The interest of Glauber in studying such a kind of system stems from fundamental issues concerning quantum mechanics \cite{Glauber0,Glauber}.
Precisely, he proposed the inverted oscillator as a toy-system through which to investigate and to explain exponential decaying of physical quantities and irreversible processes such as wavefunction collapse.
It is well known indeed that such observed phenomena cannot be deduced from first principles at the basis of quantum mechanics.
Rather, they often qualitatively emerge as possible result of some phenomenological coupling with the environment.
To circumvent such a phenomenological approach, Glauber strategically introduces the inverted quantum harmonic oscillator avoiding in this way the consideration of a disturbing bath.
The merit of such addressing is that it leads to irreversibility as a by-product of intrinsic dynamical processes.

It is worth noticing, moreover, that a measurement act is an amplification process, characterized by a strong irreversibility.
Thus Glauber's idea of an inverted quantum harmonic oscillator behaving as amplifier, appears strictly connected with the peculiar aspect of any measurement process in quantum mechanics.
Quoting Glauber \cite{Glauber}, ``\textit{it is the quantum mechanical nature of the
amplification process ultimately that makes it both noisy and irreversible}, preventing in this way strange undesired phenomena.

The inverted harmonic oscillator, of course, is an ideal system currently impossible to be experimentally realized.
However, it should be emphasized that the ``interaction engineering'' enables today the realization in laboratory of a growing number of Hamiltonian models useful to simulate the behaviour of quantum systems of interest in many contexts.
In this respect, a promising possibility is the development of sophisticated quantum circuit techniques \cite{Wang}. 
Moreover, it is important to stress that what we are interested in are physical systems that, under specific conditions, behave in such a way that they can be mathematically described by a quantum Glauber amplifier interacting with a bath of quantum harmonic oscillators.
Two examples of such systems are: 1) a single atom with huge angular momentum $J$ and subjected to a magnetic field; 2) a set of $N$ two-level atoms identically coupled to the same field.
When the systems start from an eigenstate of $J_z$ (in case of $N$ spin-1/2 the total component $J_z=\sum_ij_i^z$) with value not to far from $J$, their superfluorescent emission dynamics can be well approximated with that of the Glauber system.
For the sake of precision, such systems are non-linear amplifiers since the acceleration of the radiation rate do not continue indefinitely.
The dynamical regime related to large values of $J_z$ can be, however, quite accurately described in terms of linear quantum amplifiers \cite{Glauber0,Glauber}.

It is worth to recalling that the first experimental implementation of a Glauber-like system has been realized in a non-linear optics context through shock wave generation \cite{Gentilini}.
In this case we speak of Glauber-like amplifier since the system is properly described by the Hamiltonian of a reversed quantum harmonic oscillator rather than an inverted one: it is characterized by positive kinetic energy and a neative potential energy (in the Glauber amplifier, instead, both the energy contributions are negative).
The Glauber amplifier presents thus a high potentiality both to stimulate innovative ideas to test fundamental physical issues and to designing new devises \cite{Gentilini}.

At the light of these suggestions, in this paper we study the dynamics of a standard quantum oscillator coupled with a quantum Glauber amplifier when the Hamiltonian parameters are time-dependent.
We are interested in the interaction terms conserving the number of excitations, rather than the energy of the system.
Through the Jordan-Schwinger map, the oscillator-amplifier dynamical problem, within each dynamically invariant Hilbert space related to a precise excitation number $N$, is reduced into that of a single spin of value $N/2$.
In this way, we are able to formally construct the time evolution operator and get the exact dynamics of the system for specific initial conditions under prescribed time-dependent scenarios, such as the Rabi \cite{Rabi} and the Landau-Majorana-St\"uckelberg-Zener (LMSZ) \cite{LMSZ} ones.
We calculate the mean value of the energy when the system is initially prepared in the generalized NOON state $(\cos(\theta)\ket{N0}+e^{i\phi}\sin(\theta)\ket{0N})/\sqrt{2}$.
Furthermore, following the same spirit of the Glauber's work, we compare the dynamics of the quantum oscillator-amplifier system with that of two interacting standard quantum oscillators described by the analogous time-dependent Hamiltonian model preserving the total number of excitations.
Indeed, also in this case we are able to explicitly write the time evolution operator when precise time-dependent scenarios are considered.
Remarkable differences between the two dynamical systems are brought to light by studying the transition probability between the states $\ket{10}$ and $\ket{01}$ and the mean value of the energy for the NOON state $(\ket{10}+\ket{01})/\sqrt{2}$.

The paper is organized as follows.
In Sec. \ref{Model} the quantum oscillator-amplifier Hamiltonian model is presented together with the formal solution of the time evolution operator based on the Jordan-Schwinger map.
The mean value energy for generalized NOON states is calculated in Sec. \ref{NOON}.
Section \ref{TD Scen} reports, instead, the exact dynamics for such states is reported for two time-dependent scenarios: the Rabi \cite{Rabi} and the LMSZ \cite{LMSZ} ones.
The comparison with the exact dynamics of the system of two interacting standard harmonic oscillators is developed in Sec. \ref{Comparison}.
Finally, in the last section \ref{Conclusions} conclusive remarks are discussed.

\section{Hamiltonian Model and Formal Solution of the Dynamical Problem}\label{Model}

Let us consider the following Hamiltonian model representing a quantum optical system comprising a quantum oscillator interacting with a Glauber amplifier, namely:
\begin{equation}\label{Hho}
H_{ho}(t)={\Omega(t) \over 2}(\hat{\alpha}^\dagger \hat{\alpha}-\hat{\beta}^\dagger \hat{\beta})+\omega(t) \hat{\alpha}^\dagger \hat{\beta}+\omega^*(t) \hat{\beta}^\dagger \hat{\alpha},
\end{equation}
Systems of parametric oscillators were studied in Ref. \cite{Dod-Man} exploiting a mathematical approach based on the integrals of motion linear in the position and momentum.
The problem of oscillator equilibrium states was also discussed in Ref. \cite{Glauber-Manko,Glauber-Manko-LebedevProc}.

It can be verified that $\hat{\mathcal{N}}=\hat{\alpha}^\dagger \hat{\alpha}+\hat{\beta}^\dagger \hat{\beta}$ is constant of motion of $H_{ho}(t)$ with integer eigenvalues $N=n_1+n_2=1,2,3,\dots$.
The infinite Hilbert space can, thus, be subdivided into finite Hilbert subspaces labelled by $N$ and having dimension $N+1$.
It is worth pointing out that, on the basis of the Jordan-Schwinger map (see \cite{Jordan,Schwinger,Lemesh}):
\begin{equation}\label{JW Map}
\hat{S}_+=\hat{\alpha}^\dagger\hat{\beta}, \quad \hat{S}_-=\hat{\beta}^\dagger\hat{\alpha}, \quad \hat{S}_z={1 \over 2}(\hat{\alpha}^\dagger\hat{\alpha}-\hat{\beta}^\dagger\hat{\beta}),
\end{equation}
the effective Hamiltonian governing the dynamics of this quantum optical system can be mapped in each dynamically invariant subspace into that of a spin $s$, namely
\begin{equation}\label{Hs}
H_s(t)=\omega(t)\hat{S}_+ + \omega^*(t)\hat{S}_- + \Omega(t)\hat{S}_z,
\end{equation}
where the value of the spin is linked to the number of total excitations by $s=N/2$.
This is the key-point which allows us to derive the exact analytical expression of the time evolution operator of the two coupled quantum harmonic oscillator model.

We know \cite{Weissbluth,Hioe} that the time evolution operator $U_{1/2}$ of $H_{1/2}$ for a spin 1/2 may be written as
\begin{equation}
U_{1/2}=
\begin{pmatrix}
a(t) & b(t) \\
-b^*(t) & a^*(t)
\end{pmatrix}
\end{equation}
where $a\equiv a(t)$ and $b\equiv b(t)$ are two parameter time-functions, being solutions of the system
\begin{equation}\label{Dyn Prob}
  \left\{
  \begin{aligned}
  &\dot{a}(t)=-i{\Omega(t) \over 2} a+i\omega(t) b^{*}(t),\\
  &\dot{b}(t)=-i\omega(t) a^{*}(t)-i{\Omega(t) \over 2} b(t),\\
  &a(0)=1,\quad b(0)=0,
  \end{aligned}
  \right.
  \end{equation}
stemming directly from the equation $i\dot{U}_{1/2}=H_{1/2}U_{1/2}$ ($\hbar=1$).

The time evolution operator $U_{s}$, solution of the equation $i\dot{U}_{s}=H_{s}U_{s}$, in the standard ordered basis of the eigenstates of the third component ($\hat{S}^z$) of the spin $s$: $\{ \ket{m}, m=s,s-1, \dots, -s \}$, may be written in terms of the same two parameter time-functions $a$ and $b$ as follows \cite{Weissbluth,Hioe}
\begin{equation}\label{Ujmm'}
U_{s}^{m,m'}(a,b)\equiv \average{m|U_{s}|m'}=\sum_\mu C_{s}^{m,m'} a^{s+m'-\mu}(a^*)^{s-m-\mu}b^{m-m'+\mu}(b^*)^\mu,
\end{equation}
where \cite{Weissbluth,Hioe}
\begin{equation}\label{Cjmm'}
C_{s}^{m,m'} = (-1)^\mu {\sqrt{(s+m)!(s-m)!(s+m')!(s-m')!} \over \mu!(s+m'-\mu)!(s-m-\mu)!(m-m'+\mu)!}.
\end{equation}
Whatever $m$ and $m'$ are, the summation, formally a series generated by $\mu$ running over the integer set $\mathbb{Z}$, is a finite sum, generated by all the values of $\mu$ satisfying the condition $\text{Max}[0,m'-m]\leq\mu\leq\text{Min}[s+m',s-m]$.
In this manner, $\bigl|U_{j}^{m,m'}(a,b)\bigr|^2$ represents the probability to find the $N$-level system in the state with $z$-projection $m$ when it is initially prepared in the state with $z$-projection $m'$.
This means that by solving the problem for a single spin-1/2 we may derive and construct the solution for the analogous problem of a generic spin $s$ subjected to the same time-dependent magnetic field.

We noticed before that the total Hilbert space $\mathcal{H}$ of the two quantum harmonic oscillators is divided into dynamically invariant and orthogonal Hilbert subspaces $\mathcal{H}_{N}$ related to the different integer eigenvalues of the integral of motion $\hat{\mathcal{N}}$, that is the different values of collective excitations of the system.
We may write so
\begin{equation}
\mathcal{H}=\bigoplus_{N}\mathcal{H}_{N},
\end{equation}
with $N=n_1+n_2$, and consequently the time evolution operator $V$ of $H_{ho}$ may be cast in the following form
\begin{equation}
V=\bigoplus_{N}V_{N},
\end{equation}
where $V_{N}$ is a unitary operator responsible of the time evolution of the two harmonic oscillators in the subspace with $N$ excitations.
In this manner it is easy to see that $V$ possesses the property
\begin{equation}
\average{n_1,n_2|V|m_1,m_2}=
\left\{
\begin{aligned}
\neq 0, \qquad n_1+n_2=m_1+m_2 \\
0, \qquad n_1+n_2 \neq m_1+m_2
\end{aligned}
\right.
\end{equation}
reflecting clearly the orthogonality between the Hilbert subspaces related to different values of total excitations.

By taking into account the following equality $V_{N}=U_{N/2}$, on the basis of the J-S mapping, it is easy to check that the general probability amplitude in the coordinate representation, result
\begin{equation}\label{Prob Ampl V in coord repr}
\begin{aligned}
\average{x_1',x_2'|V|x_1,x_2} &= \sum_{n_1,n_2,m_1,m_2=0}^\infty \average{x_1',x_2'|n_1, n_2} \average{n_1,n_2|V_{N}|m_1,m_2} \average{m_1,m_2|x_1,x_2}\\
&= \sum_{N=0}^\infty \quad \sum_{n,m=0}^N \average{x_1',x_2'|n, N-n} \average{n,N-n|U_{N/2}|m,N-m} \average{m,N-m|x_1,x_2},
\end{aligned}
\end{equation}
where we used the completeness relation $\sum_{n_1,n_2=0}^\infty \ket{n_1,n_2}\bra{n_1,n_2}=1$, even representable as $\sum_{N=0}^\infty\sum_{n=0}^N \ket{n,N-n}\bra{n,N-n}=1$, with $n_1+n_2=N$.
We see that the final expression in \eqref{Prob Ampl V in coord repr} is well defined since the general term $\average{n,N-n|U^{(N/2)}|m,N-m}$ may be recovered by Eqs. \eqref{Ujmm'} and \eqref{Cjmm'}, while from the basic books of quantum mechanics it is well known that \cite{Sakurai}
\begin{equation}
\average{x|n}={1 \over \pi^{1/4}\sqrt{2^n n!}}{1\over x_0^{n+1/2}}\left( x-x_0^2{d \over dx} \right)^n \exp\left\{ -{1\over 2}\left( {x \over x_0} \right)^2 \right\},
\end{equation}
with $x_0=\sqrt{\hbar/m\tilde{\omega}}$, where $m$ and $\tilde{\omega}$ are the mass and the angular frequency of the classical oscillator, respectively.
However, it is important to point out that, though we may write the formal expression of $\average{x_1',x_2'|V|x_1,x_2}$, such a formula cannot be practically exploited since in such a case an infinite number of invariant subspace are involved; the same happens, e.g., for coherent states.

\section{Time Evolution and Energy Mean Value for NOON States}\label{NOON}

Our analysis reveals its usefulness when initial conditions involving a finite number of subspaces are considered.
In this respect, let us study the generalized NOON states
\begin{equation}\label{Gen NOON states}
\ket{\Psi_N^{\theta,\phi}(0)}= \cos(\theta)\ket{N0}+e^{i\phi}\sin(\theta)\ket{0N} , \qquad N=1,2, \dots
\end{equation}
belonging to the subspace labelled by $N$.
On the basis of our previous analysis, it is easy to see that the evolved state of the general NOON state can be formally written as $\ket{\Psi_N^{\theta,\phi}(t)}=V_N\ket{\Psi_N^{\theta,\phi}(0)}=U_{N/2}\ket{\Psi_N^{\theta,\phi}(0)}$.

It is possible to persuade oneself that, for a general excitation number $N$, we have
\begin{subequations}
\begin{eqnarray}
&&\average{\Psi_N^{\theta,\phi}(t)|{\hat{\alpha}^\dagger \hat{\alpha}-\hat{\beta}^\dagger \hat{\beta} \over 2}|\Psi_N^{\theta,\phi}(t)} \nonumber \\
&&={N \over 2}(|a|^2-|b|^2)\cos(2\theta)+\text{Re}[ab^*e^{-i\phi}] \sin(2\theta) \delta_{1N}, \\ \nonumber \\
&&\average{\Psi_N^{\theta,\phi}(t)|\hat{\alpha}\hat{\beta}^\dagger|\Psi_N^{\theta,\phi}(t)}=
[\average{\Psi_N^{\theta,\phi}(t)|\hat{\alpha}^\dagger\hat{\beta}|\Psi_N^{\theta,\phi}(t)}]^\dagger \nonumber \\
&&=-{N} ab\cos(2\theta)+{a^2e^{-i\phi}-b^2e^{i\phi} \over 2}\sin(2\theta)\delta_{1N}.
\end{eqnarray}
\end{subequations}
From the previous expression it is easy to check that for $N \geq 2$ and $\theta=\pi/4$ the two expressions vanish.
It is worth pointing out that such a circumstance is independent of the specific time-dependence of the Hamiltonian parameters.
This fact means that, when the initial condition is $\Psi_1^{\pi/4,\phi}(0)$, the time evolution of the mean value of the energy reads
\begin{equation}\label{MVE}
\average{\Psi_1^{{\pi / 4},\phi}(t)|H(t)|\Psi_1^{{\pi / 4},\phi}(t)}= {\Omega}\text{Re}[ab^*e^{-i\phi}]+\text{Re}[\omega^*(a^2e^{-i\phi}-b^2e^{i\phi})],
\end{equation}
while the following classes of NOON states
\begin{equation}
\ket{\Psi_N^{{\pi / 4},\phi}(0)}= { \ket{N0}+e^{i\phi}\ket{0N} \over \sqrt{2} }, \qquad N \geq 2,
\end{equation}
whatever the time-dependent scenario is, exhibit a constant vanishing mean value of the energy in time, that is:
\begin{equation}
\average{\Psi_N^{{\pi / 4},\phi}(t)|H(t)|\Psi_N^{{\pi / 4},\phi}(t)}=0,
\end{equation}
where $\ket{\Psi_N^{{\pi / 4},\phi}(t)}=V_{N}\ket{\Psi_N^{{\pi / 4},\phi}(0)}=U_{N/2}\ket{\Psi_N^{{\pi / 4},\phi}(0)}$.
The origin of such a result may be understood in terms of the concurrence of different factors: the symmetry of the states, the su(2) symmetry of the dynamics and the specific operators we have taken into account.
Indeed, it is possible to verify that if we consider, in the case $N=2$, the state $(\ket{20}+\ket{11}+\ket{02})/\sqrt{3}$, we get a non-vanishing mean value of the energy.
Analogously, if consider the non-linear operators $(\alpha\beta^\dagger)^2$, $(\alpha^\dagger\beta)^2$, $(\hat{\alpha}^\dagger \hat{\alpha}-\hat{\beta}^\dagger \hat{\beta})^2/4$ and the initial state $\ket{\Psi_2^{\theta,\phi}(0)}$ we obtain
\begin{subequations}
\begin{eqnarray}
&&\average{\Psi_2^{\theta,\phi}(t)|{(\hat{\alpha}^\dagger \hat{\alpha}-\hat{\beta}^\dagger \hat{\beta})^2 \over 4}|\Psi_2^{\theta,\phi}(t)} \nonumber \\
&&=|a|^4+|b|^4+2\text{Re}[(ab^*)^2e^{-i\phi}]\sin(2\theta), \\ \nonumber \\
&&\average{\Psi_2^{\theta,\phi}(t)|(\hat{\alpha}\hat{\beta}^\dagger)^2|\Psi_2^{\theta,\phi}(t)}=
[\average{\Psi_2^{\theta,\phi}(t)|(\hat{\alpha}^\dagger\hat{\beta})^2|\Psi_2^{\theta,\phi}(t)}]^\dagger \nonumber \\
&&={(a^4e^{-i\phi}+b^4e^{i\phi})\cos(\theta)\sin(\theta)+a^2 b^2 \over 2},
\end{eqnarray}
\end{subequations}
which are different from zero also for $\theta=\pi/4$, so that the mean value of the energy is neither vanishing nor constant in time.
We stress, moreover, that such a calculation shows that correlations between the inverted and the normal quantum harmonic oscillator are present since the covariances of the operators under scrutiny do not vanish.

\section{Time evolution under specific scenarios}\label{TD Scen}

In this section we analyse specific time-dependent scenarios to show the practical applicability of our analysis and results previously discussed.

\subsection{Time-Independent Case}

\begin{figure}[htp]
\centering
{\includegraphics[scale=.8]{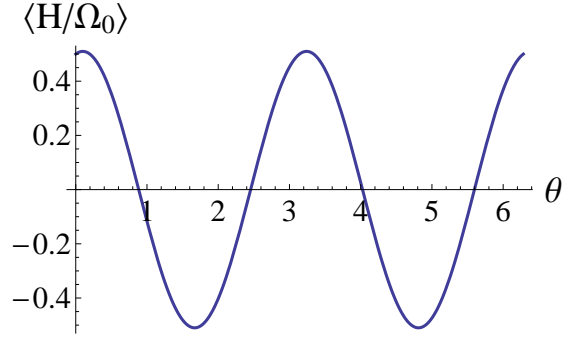} }
\captionsetup{justification=justified,format=plain,skip=4pt}%
\caption{(Color online) Mean value of the energy in Eq. \eqref{MVE} scaled with respect to the parameter $\Omega(t)=\Omega_0$, with $\omega(t)=0.1\Omega_0$ and versus the parameter $\theta$, when the quantum oscillator-amplifier system starts from the state $\ket{\Psi_1^{\theta,0}}$ [Eq. \eqref{Gen NOON states}].}\label{fig:EMVConstG}
\end{figure}

First of all, let us take into account the simplest case, that is when the Hamiltonian parameters are time-independent: $\Omega(t)=\Omega_0$ and $\omega(t)=\omega_0$.
Moreover, let us consider, for simplicity, $\omega_0$ a real parameter; such a choice is justified by the fact that a unitary transformation (a rotation with respect to $\hat{z}$) can be always performed in order to make $\omega$ a real parameter.
In this instance the two time-function parameters $a$ and $b$, solving the system in Eq. \eqref{Dyn Prob}, acquire the following form
\begin{equation}
a(t) = \left[ \cos(\tau) - i {\Omega_0 \over 2\hbar\nu} \sin(\nu t) \right], \quad
b(t) = -i {\omega_0 \over \hbar\nu} \sin(\nu t),
\end{equation}
with $\hbar\nu \equiv \sqrt{\Omega_0^2/4+\omega_0^2}$.
In this way we can get explicit analytical expressions for all the formulas we obtained before.
We can calculate, for example, the time evolution of the mean value of the energy.
In Fig. \ref{fig:EMVConstG} we report the $\theta$-dependence of such a quantity when the system is initialized in the state $\ket{\Psi_1^{\theta,0}}$, whose general expression is reported in Eq. \eqref{MVE};it is easy to see that in this time scenario, as expected, the mean value energy is constant in time and depends only on the parameter $\theta$ [see Eq. \eqref{Gen NOON states}].

\subsection{Rabi Scenario}

Now, we consider the real coupling parameter oscillating in time, namely $\omega(t)=\omega_0\cos(\nu_0t)$, and leave the parameter $\Omega(t)=\Omega_0$ constant.
Such a physical scenario, in terms of the spin language, may be reduced to the well known Rabi model \cite{Rabi}.
Precisely, under the conditions $\omega_0/\Omega_0 \ll 1$ and $\nu_0=\Omega_0/2\hbar$ (resonance condition), only the rotating terms of the time-dependent transverse field ($\omega$) are relevant for the dynamics of the system, so that the counter rotating ones can be disregarded.
In this instance, the coupling parameter becomes $\omega=\cos(\nu_0t)-i\sin(\nu_0t)$.
The related dynamical problem may be exactly solved and the expressions of $a$ and $b$ defining the time-evolution operator, solutions of \eqref{Dyn Prob} read
\begin{equation}
a=\cos(k\tau')e^{-i\tau'}, \quad b=-i\sin(k\tau')e^{-i\tau'}, \qquad \tau'=\nu_0t, \quad k=\omega_0/\Omega_0.
\end{equation}
The time evolution of the mean value of the energy [Eq. \eqref{MVE}] when the quantum oscillator-amplifier system starts from the state $\ket{\Psi_1^{\pi/4,0}}$ [Eq. \eqref{Gen NOON states}], is reported in Fig. \ref{fig:EMVRG}, for $\omega_0/\Omega_0=0.1$, with respect to the dimensionless time $\tau'=\nu_0t$.
We see the presence of the typical oscillatory regime of the Rabi scenario, being
\begin{equation}
\average{\Psi_1^{\pi/4,0}|U^\dagger H U|\Psi_1^{\pi/4,0}}=\omega_0\cos(k\tau').
\end{equation} 

\subsection{Landau-Majorana-St\"uckelberg-Zener Scenario}

The Landau-Majorana-St\"uckelberg-Zener (LMSZ) scenario \cite{LMSZ} is characterized by a linear longitudinal (in the $z$ direction) ramp, namely, $\Omega(t)=\gamma~t$, with $t \in (-\infty,\infty)$ and a transverse (along the $x$ direction) constant field, $\omega=\omega^*=\omega_0$.
The LMSZ scenario is an ideal model since it provides for an infinite duration of the physical procedure.
To comply with more physical experimental condition, it is more appropriate to consider finite values for the initial and final time instants.
In this case, the exact solution of $a(t)$ and $b(t)$ for the system in Eq. \eqref{Dyn Prob} read \cite{Vit-Garr}
\begin{subequations}\label{Exact a b LMSZ}
\begin{align}
a=&{\Gamma_f(1-i\chi) \over \sqrt{2\pi}} \times \nonumber \\
&\left[ D_{i\chi}(\sqrt{2}e^{-i\pi/4}\tau) ~ D_{-1+i\chi}(\sqrt{2}e^{i3\pi/4}\tau_i)
+D_{i\chi}(\sqrt{2}e^{i3\pi/4}\tau) ~ D_{-1+i\chi}(\sqrt{2}e^{-i\pi/4}\tau_i) \right],
\\ \nonumber\\
b=&{\Gamma_f(1-i\chi) \over \sqrt{2\pi\chi}} e^{i\pi/4} \times \nonumber \\
&\left[ -D_{i\chi}(\sqrt{2}e^{-i\pi/4}\tau) ~ D_{-1+i\chi}(\sqrt{2}e^{i3\pi/4}\tau_i)
+D_{i\chi}(\sqrt{2}e^{i3\pi/4}\tau) ~ D_{-1+i\chi}(\sqrt{2}e^{-i\pi/4}\tau_i) \right]
\end{align}
\end{subequations}
where $\chi=2\omega_0^2/\hbar\gamma$ is the LMSZ parameter, $\Gamma_f$ is the gamma function, $D_\nu(z)$ are the parabolic cylinder functions \cite{Abramowitz} and $\tau=\sqrt{\gamma/\hbar}~t$ is a time dimensionless parameter; $\tau_i$ identify the initial time instant.

\begin{figure}[htp]
\centering
\subfloat[][]{\includegraphics[scale=.6]{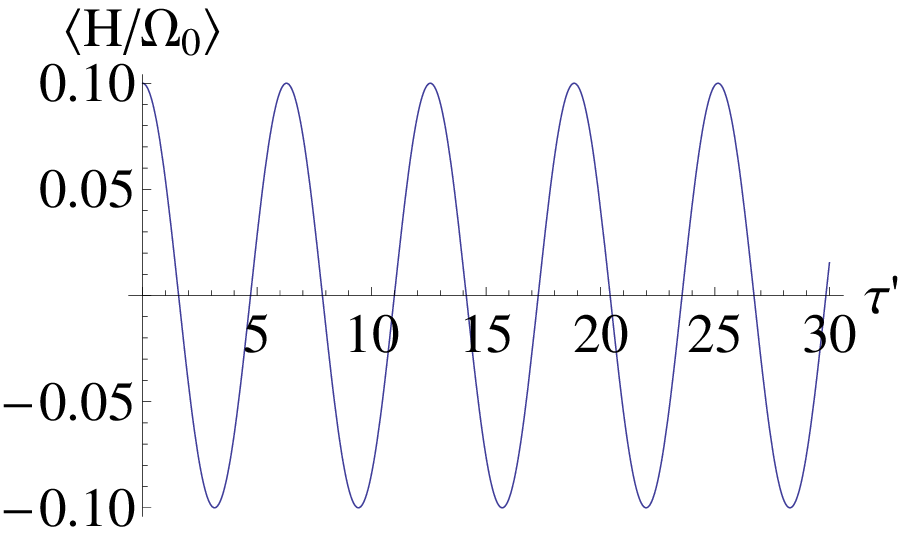}\label{fig:EMVRG}}
\qquad
\subfloat[][]{\includegraphics[scale=.6]{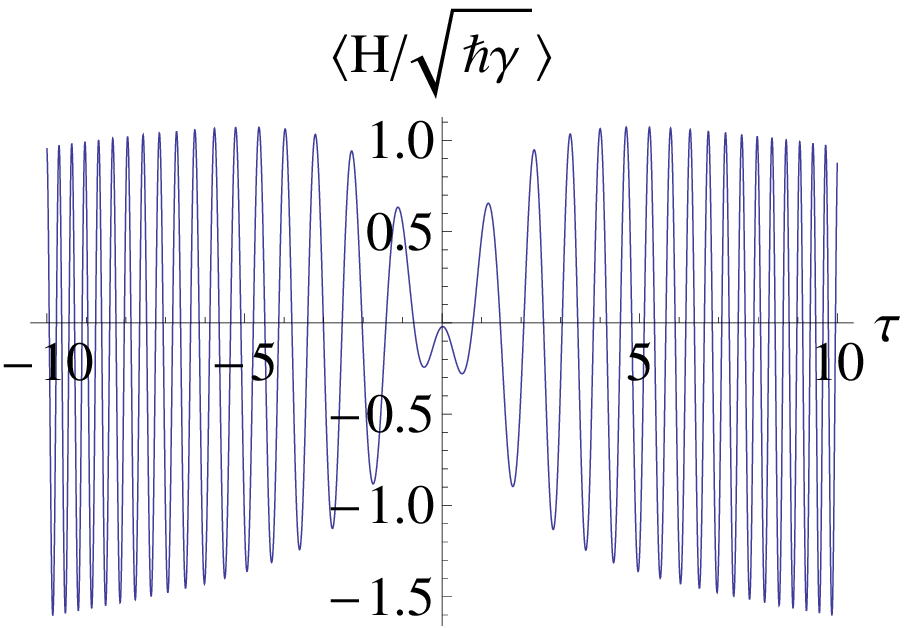}\label{fig:EMVLZG}}
\\
\subfloat[][]{\includegraphics[scale=.6]{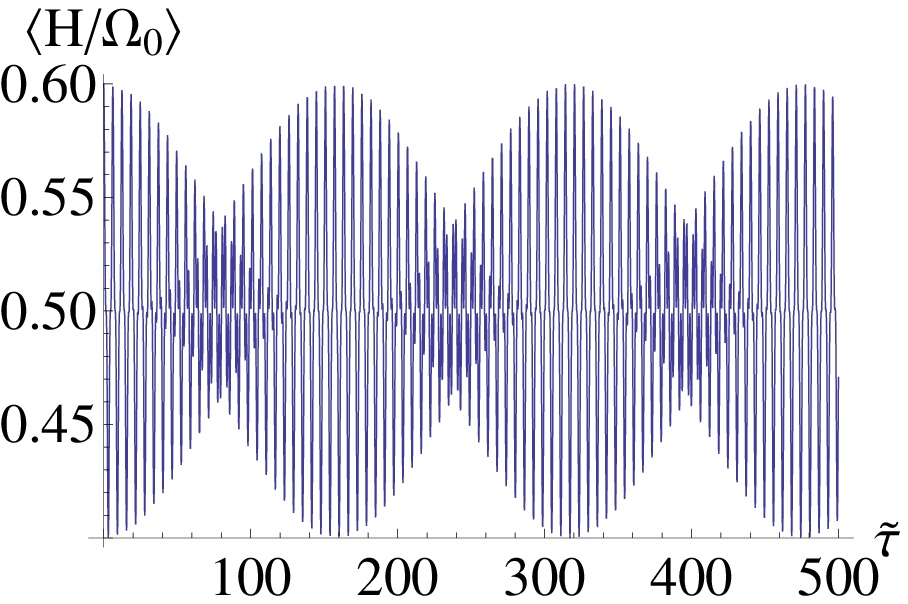}\label{fig:EMVRST}}
\qquad
\subfloat[][]{\includegraphics[scale=.6]{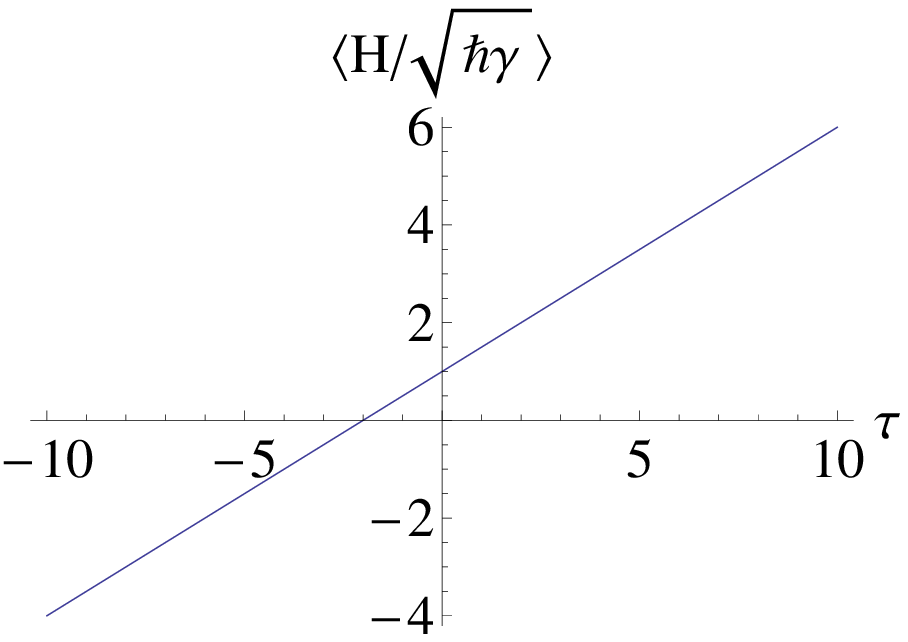}\label{fig:EMVLZST}}
\captionsetup{justification=justified,format=plain,skip=4pt}%
\caption{(Color online) Mean value of the energy when the oscillator-amplifier [oscillator-oscillator] system starts from $\ket{\Psi_1^{\pi/4,0}}$ for a) the Rabi scenario with $\omega_0/\Omega_0=0.1$ and b) the LMSZ scenario when $\omega_0^2/\hbar\gamma=1$. Mean value of the energy when the oscillator-oscillator system starts from $\ket{\Psi_1^{\pi/4,0}}$ for c) the Rabi scenario with $\omega_0/\Omega_0=\omega_0/2\hbar\nu_0=0.1$ and d) the LMSZ scenario when $\omega_0^2/\hbar\gamma=1$.}
\end{figure}

The plot of the mean value of the energy for the initial condition $\ket{\Psi_1^{\pi/4,0}}$ in such a scenario is reported in Fig. \ref{fig:EMVLZG}.
We note that the curve is symmetric with respect to the time instant ($t=0$) in which the avoided crossing occurs.
This circumstance can be understood by writing the state of the system at a general time instant $t$:
\begin{equation}
U_{1/2}(t)\ket{\Psi_1^{\pi/4,0}}={(a+b)\ket{10}+(a^+-b^*)\ket{01} \over \sqrt{2}},
\end{equation}
and by considering that under the LMSZ scenario $a(t)$ $[b(t)]$ goes from $1$ $[0]$ to $0$ $[1]$.
Thus, it means that the system reaches asymptotically the state $(\ket{10}-\ket{01}) / \sqrt{2}$ which differs from the initial condition $[(\ket{10}+\ket{01}) / \sqrt{2}]$ only for the relative phase factor.

\section{Comparison with the two interacting standard harmonic oscillator model}\label{Comparison}

Let us consider now the same model for two standard quantum harmonic oscillators:
\begin{equation}\label{Hho St}
H_{ho}'={\Omega(t) \over 2}(\hat{\alpha}^\dagger \hat{\alpha}+\hat{\beta}^\dagger \hat{\beta})+\omega(t) \hat{\alpha}^\dagger \hat{\beta}+\omega^*(t) \hat{\beta}^\dagger \hat{\alpha},
\end{equation}
It is easy to understand that the total excitation number $\hat{\mathcal{N}}=\hat{\alpha}^\dagger \hat{\alpha}+\hat{\beta}^\dagger \hat{\beta}$ is a constant of motion for this Hamiltonian too, $[H'(t),\mathcal{N}]=0$.
This fact implies that, also this time, we have an infinite number of dynamical invariant Hilbert subspaces related to the different eigenvalues $N=1,2\dots$ of $\hat{\mathcal{N}}$.
It is possible to persuade oneself that, in this case, the two oscillator dynamical problem may be mapped within the $N+1$-dimensional subspace (linked to the eigenvalue $N$ of $\mathcal{N}$) into a spin-$N/2$ dynamical problem related to the following Hamiltonian:
\begin{equation}
H_N={N \over 2} \Omega(t) \hat{\mathbb{1}} + \omega(t) \hat{S}^+ + \omega^*(t) \hat{S}^-
\end{equation}
In this instance, the time evolution operator governing the dynamics within such a subspace can be written as
\begin{equation}
V_N(t)=\exp\left\{-{i \over \hbar}{N \over 2}\int_0^t\Omega(t')dt'\right\}U_{N/2}(t)
\end{equation}
with $U_{N/2}$ defined in Eq. \eqref{Ujmm'}, where $a(t)$ and $b(t)$ are the solutions of the system of differential equations originating from the spin-1/2 dynamical problem:
\begin{equation}\label{Dyn Prob st}
  \left\{
  \begin{aligned}
  &\dot{a}(t)=i\omega(t) b^{*}(t),\\
  &\dot{b}(t)=-i\omega(t) a^{*}(t),\\
  &a(0)=1,\quad b(0)=0,
  \end{aligned}
  \right.
\end{equation}

In the time-independent case, $\Omega(t)=\Omega_0$ and $\omega(t)=\omega_0$, the expressions of $a(t)$ and $b(t)$ are
\begin{equation}\label{a b const ho st}
a(t) = \cos(\omega_0 ~ t/\hbar), \quad
b(t) = -i  \sin(\omega_0 ~ t/\hbar).
\end{equation}
In the Rabi scenario, that is, when $\omega(t)=\omega_0e^{-i\nu_0t}$, the two parameter time functions read instead
\begin{equation}
a(t) = \cos(\tilde{\tau}), \quad
b(t) = -i {\omega_0 \over \hbar\nu_R} \sin(\tilde{\tau}), \qquad
\tilde{\tau}=\nu_R ~ t, \quad
\nu_R=\sqrt{\nu_0^2+\omega_0^2/\hbar^2}.
\end{equation}
In the LMSZ scenario ($\Omega(t)=\gamma~t$, $\omega=\omega^*=\omega_0$) the expressions of the two time functions are very similar to those in Eq. \eqref{a b const ho st}, namely
\begin{equation}
a(t) = \cos(\sqrt{\chi/2}~\tau), \quad
b(t) = -i  \sin(\sqrt{\chi/2}~\tau), \qquad
\tau=\sqrt{\gamma/\hbar} ~ t, \quad \chi=2\omega_0^2/\hbar\gamma,
\end{equation}
since $\Omega(t)$, in case of two interacting standard oscillators plays no role in determining $a(t)$ and $b(t)$, as it is clear from Eq. \eqref{Dyn Prob st}.

The time evolution of the mean value of the energy when the two interacting quantum oscillators are initially prepared in $\ket{\Psi_1^{\pi/4,0}}$ is reported in Figs. \ref{fig:EMVRST} and \ref{fig:EMVLZST} for the Rabi and the LMSZ scenario, respectively, with $\omega_0/\Omega_0=\omega_0/2\hbar\nu_0=0.1$ in the first case and $\omega_0^2/\hbar\gamma=1$ in the second case.
We see that the Rabi scenario preserves, of course, its qualitative oscillatory regime, although the oscillation is consistently different presenting a beat effect, since
\begin{equation}
\average{\Psi_1^{\pi/4,0}|U^\dagger H U|\Psi_1^{\pi/4,0}}={\Omega_0 \over 2}+\omega_0\cos(\nu_0t)\left[\cos^2(\nu_Rt)+{\omega_0^2\over\hbar^2\nu_R^2}\sin(\nu_Rt)\right]
\end{equation}
A drastic change, instead, happens in the LMSZ scenario for which we have
\begin{equation}
\average{\Psi_1^{\pi/4,0}|U^\dagger H U|\Psi_1^{\pi/4,0}}=\sqrt{\hbar\gamma}~\tau+\omega_0.
\end{equation}
This is due to the fact that the dynamics of the two-oscillator system is unaffected by the parameter $\Omega(t)$.
The physical reason is that, in the LMSZ framework, $\Omega(t)$ is the main parameter driving the time evolution of the system and realizing the characteristic LMSZ dynamics as it happens for the oscillator-amplifier system.

To appreciate the difference between the dynamics of the two quantum systems even better, let us consider now the time evolution of the state $\ket{N0}$; it is easy to see that
\begin{equation}
P_{N0}^{0N}=\average{0N|N0(t)}\equiv\average{0N|V_{N}(t)|N0}=\average{0N|U_{N/2}(t)|N0}=|b(t)|^{2N}.
\end{equation}
We note that in the time-independent case, for the two standard oscillators, $P_{N0}^{0N}=\sin^{2N}(\omega_0t/\hbar)$ presents oscillations with maximum amplitude.
In the case of an oscillator coupled with a Glauber amplifier, instead, such a transition probability, $P_{N0}^{0N}=(\omega_0 / \hbar\nu)^{2N} \sin^{2N}(\nu t)$, cannot reach, in general, the maximum value $P_{N0}^{0N}=1$, unless in the more trivial case $\Omega_0=0$.
The opposite situation occurs in the case of the Rabi scenario.
We have, indeed, $P_{N0}^{0N}=(\omega_0 / \hbar\nu_R)^{2N} \sin^{2N}(\tilde{\tau})$ for two oscillators and $P_{N0}^{0N}=\sin^{2N}(k\tau')$ for the quantum oscillator-amplifier system.
This circumstance can be traced back to the fact that the resonant condition cannot be satisfied in the case of two standard oscillators ($\Omega_0$ plays no role in the dynamics).
\begin{figure}[htp]
\centering
\subfloat[][]{\includegraphics[scale=.6]{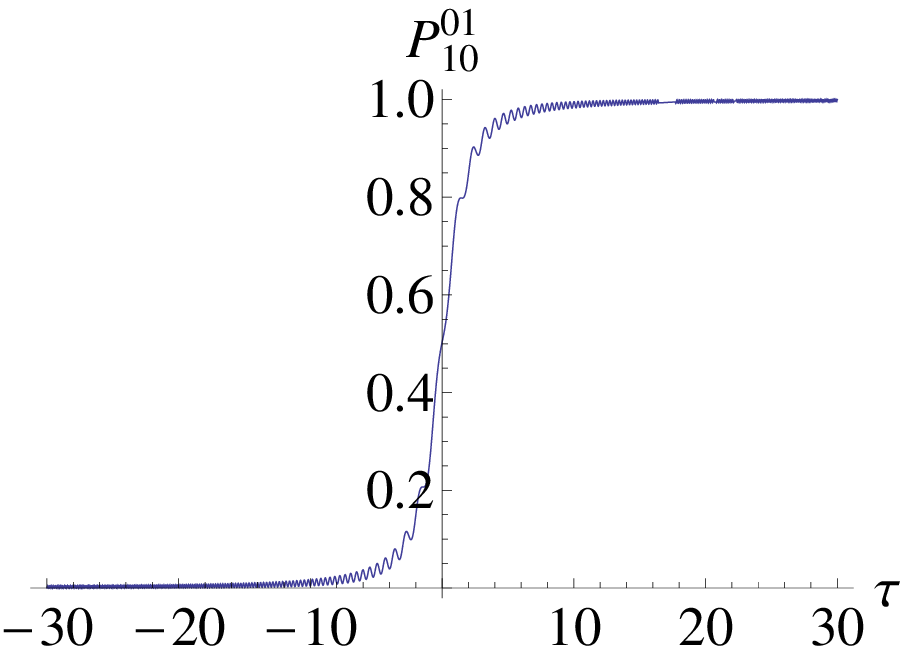}\label{fig:PG}}
\qquad
\subfloat[][]{\includegraphics[scale=.6]{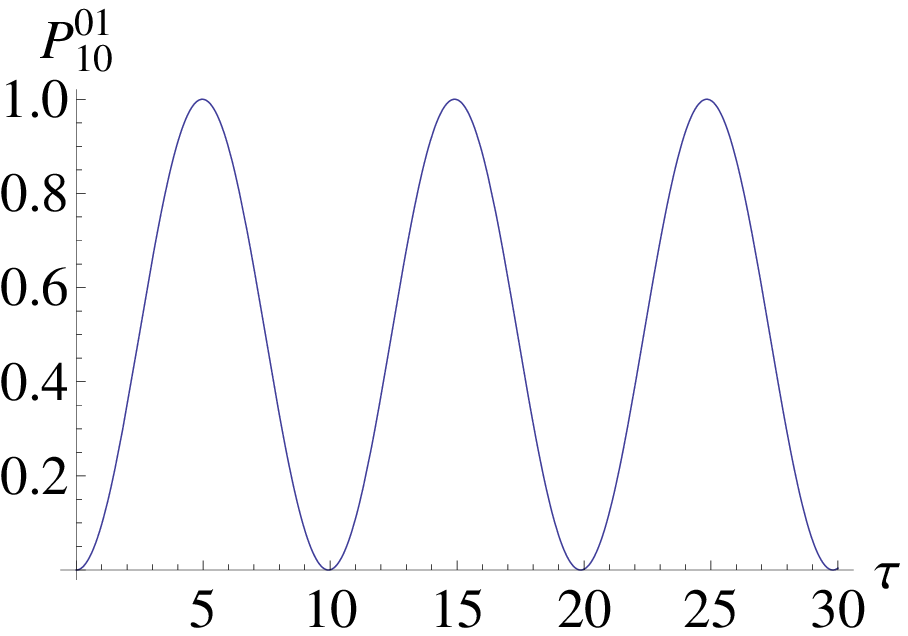}\label{fig:PST}}
\captionsetup{justification=justified,format=plain,skip=4pt}%
\caption{(Color online) Time evolution of the probability $P_{10}^{01}=\average{01|U_{1/2}(t)|10}=|b(t)|^{2}$ in the LMSZ scenario with $\omega_0^2/\hbar\gamma=1$ for a) the oscillator-amplifier system and b) the oscillator-oscillator system.}
\end{figure}
Finally, we underline that the LMSZ scenario, in case of two standard oscillators, does not generate the typical LMSZ transition probability, but the behaviour of $P_{N0}^{0N}$ results sinusoidal in time: $P_{N0}^{0N}=\sin^{2N}(\tilde{\tau})$.
The oscillator-amplifier system, instead, exhibits an asymptotic full transition from $\ket{N0}$ to $\ket{0N}$ under adiabatic conditions, that is, when $\omega_0/\gamma \ll 1$.
The two different probabilities for the LMSZ scenario are reported in Figs. \ref{fig:PG} and \ref{fig:PST} in case of $N=1$ with $\omega_0^2/\hbar\gamma=1$.

\section{Conclusive remarks}\label{Conclusions}

Jordan [9] and Schwinger [10] have shown that the angular momentum operators can be expressed in terms of quadratic expressions of two bosonic annihilation and creation operators $\hat{a}_1$, $\hat{a}_2$ and $\hat{a}_1^\dagger$, $\hat{a}_2^\dagger$.
Such a general statement is known as Jordan-Wigner map \eqref{JW Map}.
Namely, given three $N \times N$-matrices $A$, $B$ and $C$ such that $[A,B]=C$, the three operators $\hat{A}=\sum_{j,k=1}^N A_{jk} \hat{a}_j^\dagger \hat{a}_k$, $\hat{B}=\sum_{j,k=1}^N B_{jk} \hat{a}_j^\dagger \hat{a}_k$ and $\hat{C}=\sum_{j,k=1}^N C_{jk} \hat{a}_j^\dagger \hat{a}_k$ satisfy the commutation relation $[\hat{A},\hat{B}]=\hat{C}$.
If matrices $A$, $B$ and $C$ are $2 \times 2$ Pauli matrices this statement provides possibility to construct all the spin-states with $S=0,1/2,1,3/2,\dots$ in terms of two oscillator states $\ket{n_1,n_2}$, where $n_1+n_2=2s+1$.
The original idea was to exploit the solutions of the non-stationary Schr\"odinger equation related to two-mode parametric oscillators to map them into solutions of the Schr\"odinger equation related to non-stationary Hamiltonians linear in the generators of the SU$(2)$ group.

In this paper, instead, we adopted exactly the opposite strategy.
Through the Jordan-Wigner mathematical trick, within each invariant subspace, it is possible to map the dynamical problem of the oscillator-amplifier system into that of a single spin-$j$ (the value of $j$ depends on the dimension of the subspace) characterized by a Hamiltonian linear in the SU(2) generators.
Thanks to the knowledge of the formal expression of the SU(2)-group elements (representing the time evolution operators solution of the dynamical problem of the general single spin-$j$) we constructed the time evolution operator of the quantum oscillator-amplifier system.
Moreover, on the basis of the knowledge of exact solutions pertaining to specific time-dependent scenarios, we studied the exact dynamics of the oscillator system.
Following the same approach, we solved and analysed also the dynamics of two interacting (standard) quantum harmonic oscillators.
A comparison between some dynamical properties exhibited by the oscillator-amplifier system and the oscillator-oscillator one has allowed us to bring to light relevant physical analogies and differences.

We emphasize that other exact or approximated solutions of the single qubit dynamical problem \cite{Bagrov,Kuna,DasSarma,Mess-Nak,MGMN,GdCNM1} may be exploited to study the dynamics of the systems under scrutiny subjected to different physical conditions with possible useful applications.
It is worth to point out that an analogous approach has been used to treat and solve dynamical problems of interacting qubit system \cite{GMN,GMIV,GBNM} and proved to be useful to bring to light relevant physical effects \cite{GLSM,GVM1,GVM2,GIMGM}.

We wish to point out that the same strategy can be used for arbitrarily Hamiltonians presenting a linear form in the generators of any Lie algebra.
Also these generators, indeed, can be expressed in terms of bosonic or fermionic creation and annihilation operators as quadratic forms in operators with time-dependent coefficients.
In that case, it results of basic importance the knowledge of exact solutions of dynamical problems characterized by different symmetries.
In this respect, it is interesting to underline that a solution method has been recently proposed for dynamical problems related to su(1,1) Hamiltonians \cite{GdCKM,GdCNM2}.
Such kind of Hamiltonians are very useful and important to treat and study open quantum systems living in finite Hilbert spaces and described by pseudo-Hermitian Hamiltonians, such as $PT$-symmetry physical systems \cite{Ruter,Liertzer,Bittner,Schindler,Tripathi}.
Moreover, it is interesting to stress that in case of infinite dimensional Hilbert spaces, like quantum oscillators and amplifier, the representation of the SU(1,1) group results unitary and then appropriate to describe coherent dynamics of closed physical systems.

Finally, a further possible perspective of the present work could be investigating the same system in presence of a bath of quantum oscillators.
However, the correspondent more complex Hamiltonian model would be no longer characterized by the existence of invariant finite dimensional su(2)-symmetry subspaces.
Moreover, it would be very difficult to use the Jordan-Schwinger map in order to simplify the problem by describing it in terms of spin variables.
To appreciate this point it is enough to consider that a Glauber amplifier in a bath of oscillators could be described, in principle, in terms of several coupled spins; but such a problem would present analytical difficulties comparable with those appearing in the oscillator formulation.
In this instance, thus, the exact treatment of the dynamical problem would become hard, requiring, as a consequence, the consideration of other approaches as, for example the ones reported in Refs. \cite{Glauber-Manko,Glauber-Manko-LebedevProc}.
It is interesting, for example, even the approach reported in Ref. \cite{Lorenzen} based on the derivation of the Gorini-Kossakowski-Lindbland-Sudarshan equation\cite{GKLS} master equation as well as on the Wigner function to get intriguing physical dynamical feature of the system.
In our case, however, the presence of time-dependent Hamiltonian parameters gives rise to further difficulties.
But to this end, a possible approach would be the one based on the so-called quantum-classical Liouville equation stemming from the partial Wigner transpose.
In this instance, the oscillator variables are treated as classical parameters making less cumbersome the numerical analysis of the problem \cite{Sergi}.

\Ignore{
As an example in the following we consider the case of $N=1$, that is when only one excitation is present in the system, so that the two quantum harmonic oscillator time evolution is restricted in the two dimensional Hilbert subspace.
In this instance the formula \eqref{Prob Ampl V in coord repr} becomes
\begin{equation}
\average{x_1',x_2'|V|x_1,x_2}={2 \over \pi x_{01}x_{02}}\exp\left\{ -{1 \over 2}\left( {x_1^2+x_1^{'2} \over x_{01}^2}+{x_2^2+x_2^{'2} \over x_{02}^2} \right) \right\} \left[ {x_1x_1' \over x_{01}^2}a+{x_1x_2' \over x_{01}x_{02}}b-{x_1'x_2 \over x_{01}x_{02}}b^*+{x_2x_2' \over x_{02}^2}a^* \right]
\end{equation}
with $x_{0i}=\sqrt{\hbar/m_i\tilde{\omega}_i}$ ($i=1,2$).

For, through this procedure we are able to write formally the exact analytical expression of the probability amplitude, and then of the probability, to find two quantum harmonic oscillators, coupled according the model in Eq. \eqref{Hho}, when they pass from a configuration to another one in the coordinate space.
This means that, in this manner, for this interaction model we have reduced the problem of two coupled quantum harmonic oscillators into that of a single spin-1/2.
Therefore, exploiting the known exact solutions and the methods present in literature to get new analytically solvable models of single spin 1/2 \cite{KN, Das Sarma,Mess-Nak,GMN}, we may derive and solve exactly several time-dependent scenarios for two quantum harmonic oscillators coupled according the model written in Eq. \eqref{Hho}.
}

\end{document}